\documentclass[runningheads]{llncs}
\usepackage{graphicx}
\usepackage{amsmath}
\newcommand{\anatomix}{AnatoMix }

\newcommand{\nd}{$N_d$ }
\newcommand{\nt}{$N_t$ }

\DeclareMathOperator*{\argmax}{arg\,max}

\begin{document}
\title{AnatoMix: Anatomy-aware Data Augmentation for Multi-organ Segmentation}

%
%
\author{Chang Liu\inst{1} \and
Fuxin Fan\inst{1} \and
Annette Schwarz\inst{1} \and
Andreas Maier\inst{1} }

\authorrunning{C. Liu et al.}
%
\institute{Pattern Recognition Lab, Friedrich-Alexander-Universität, 91058 Erlangen, Germany}
%
\maketitle              
\begin{abstract}
Multi-organ segmentation in medical images is a widely researched task and can save much manual efforts of clinicians in daily routines. Automating the organ segmentation process using deep learning (DL) is a promising solution and state-of-the-art segmentation models are achieving promising accuracy. 
In this work, We proposed a novel data augmentation strategy for increasing the generalizibility of multi-organ segmentation datasets, namely AnatoMix. 
By object-level matching and manipulation, our method is able to generate new images with correct anatomy, i.e. organ segmentation mask, exponentially increasing the size of the segmentation dataset. 
Initial experiments have been done to investigate the segmentation performance influenced by our method on a public CT dataset.
Our augmentation method can lead to mean dice of 76.1, compared with 74.8 of the baseline method.

\end{abstract}
\section{Introduction}

Multi-organ segmentation is a widely applied clinical procedure, i.e. for treatment planning of the radiation therapy. The organs-at-risk must be delineated for planning the radiation procedure. In additional, the segmentation of the radiation sensitive organs can be applied to the estimation of CT effective dose.
Deep learning (DL) based multi-organ segmentation models have achieved promising performance and the leading models in public multi-organ segmentation challenges shows that the DL-based models can have competing performance as human annotations \cite{isensee2021nnu}. However, DL models are usually data-demanding and a generalizable DL model needs large-scale dataset for training. On the one hand large-scale organ segmentation datasets are emerging, for example AbdomenCT-1K \cite{ma2021abdomenct}, AMOS \cite{ji2022amos} and TotalSegmentator \cite{wasserthal2023totalsegmentator}. 
On the other hand, many researches have been done to expand the generalizability of the segmentation dataset while keeping the robust training. 
In this work, we proposed a novel data augmentation strategy for multi-organ segmentation dataset inspired by human anatomy, namely AnatoMix. 


Regarding the task of multi-organ segmentation, the intrinsic constraints of human anatomy must be respected and the augmentation method should be able to generate anatomically correct images.
Inspired by the object-level data augmentation methods and the intrinsic connection of human anatomy, we proposed \anatomix, a organ-level data augmentation method for multi-organ segmentation tasks. \anatomix combines the anatomical structures from different patients and thus leading to exponential increase of size of dataset. 
\anatomix also accounts for the human anatomy, i.e. the correct location and size of the organs in the generated images. 
A major difference of \anatomix is that it will not output unlimited augmented because the reasonable combination is also considered. 

\begin{figure}[t]
    \centering
    \includegraphics[width=\textwidth]{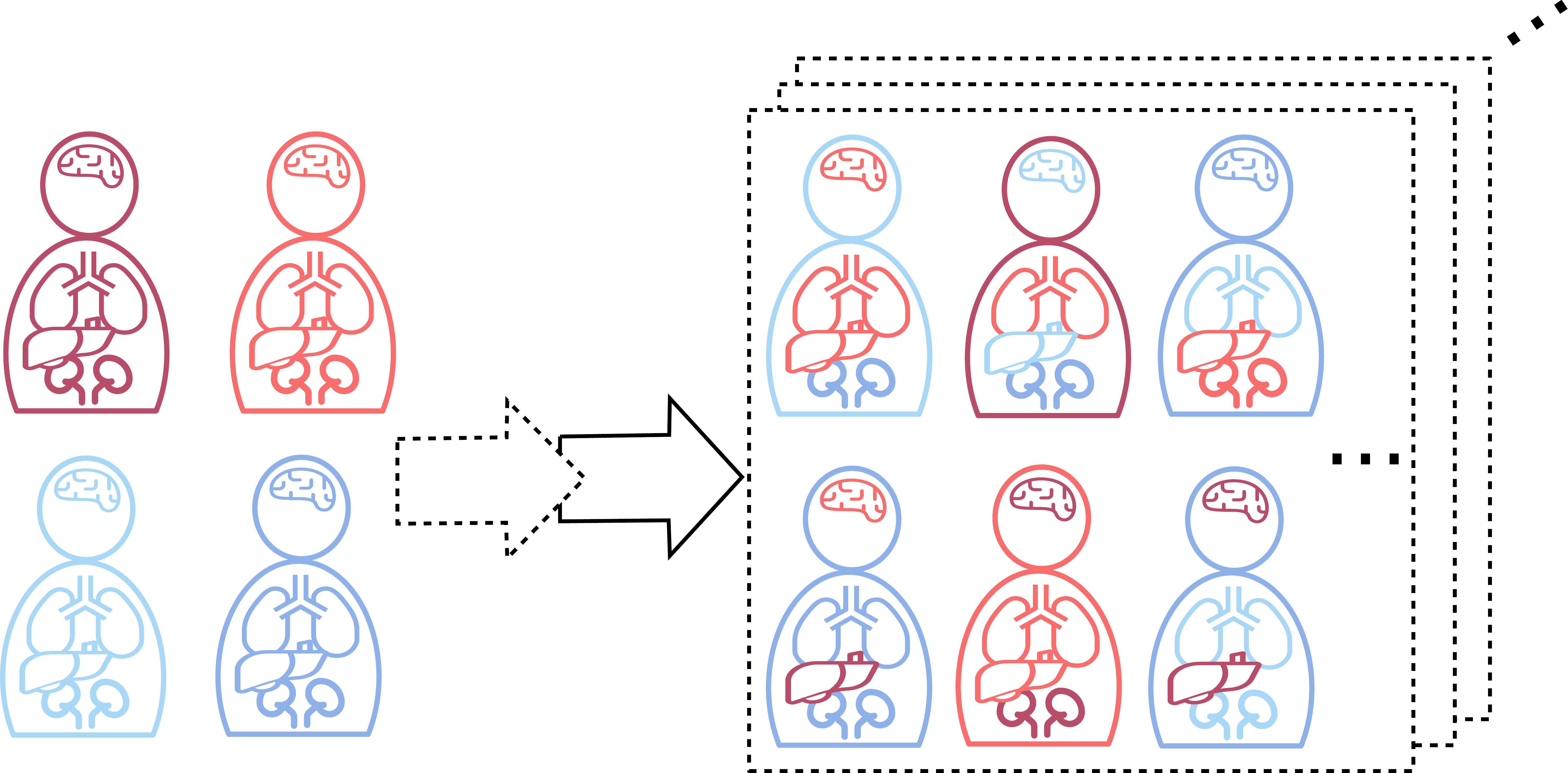}
    \caption{Illustration of the concept of \anatomix.}
    \label{fig:illustration}
\end{figure}

\section{Method}
\begin{figure}[ht]
    \centering
    \includegraphics[width=\textwidth]{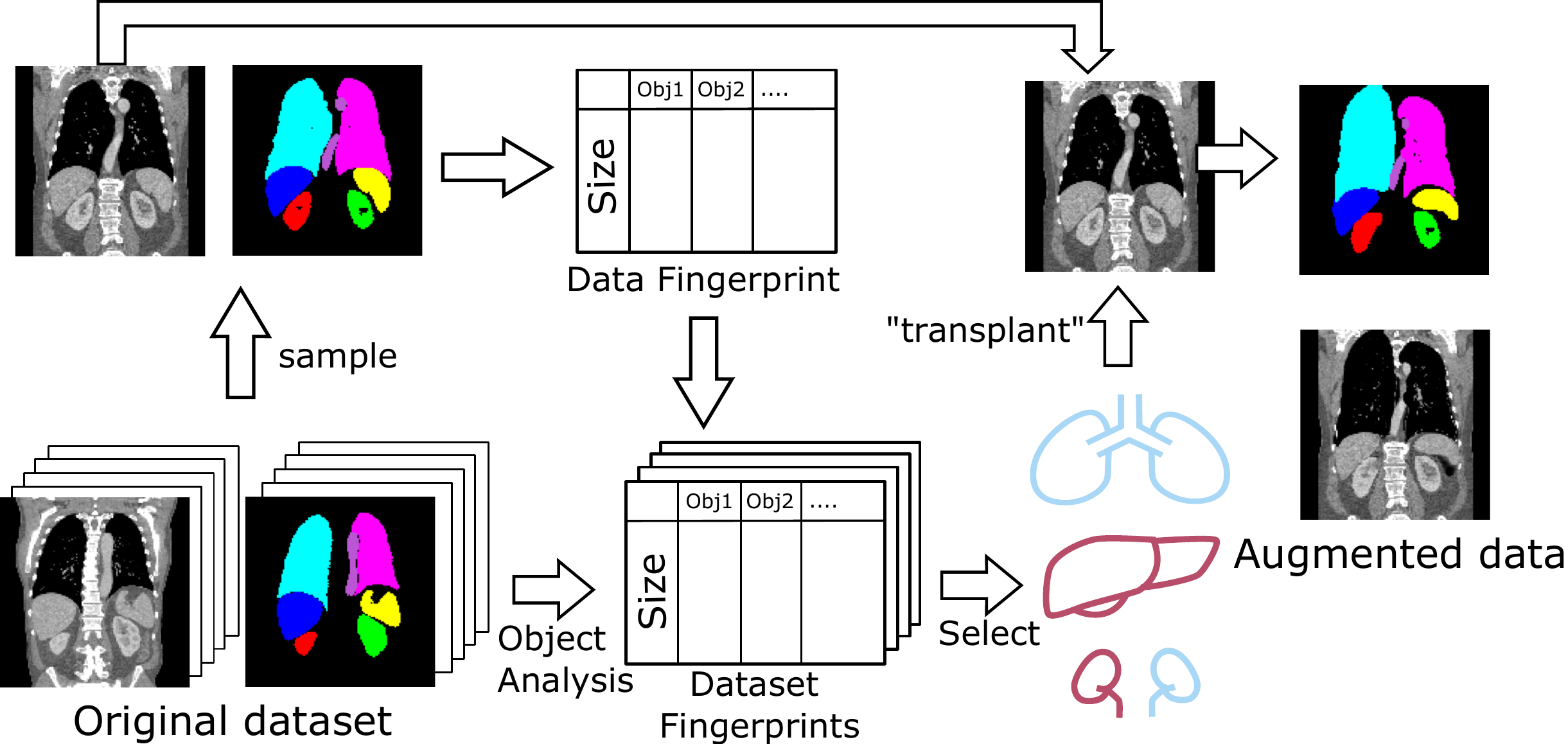}
    \caption{The detailed pipeline of \anatomix.}
    \label{fig:pipelinev2}
\end{figure}

The pipeline of \anatomix is shown in Fig. \ref{fig:pipelinev2} , \anatomix contains two major steps in pipeline: augmentation planning and object transplant.
Experiments are designed to investigate \anatomix on CT-ORG dataset.


\subsection{Augmentation Planning}
The concept of \anatomix is to separate the annotated organs from the images in the dataset, and recombine the organs onto the background image to generate new images. 
A segmentation dataset with \nd images and \nt target organs can result in $N_d^{N_t+1}$ possible combinations and each combination correspond to a new image with correct organ segmentation mask. 
In the case of the CT-ORG dataset, where 28 CT images with 4 annotated organs are used for training, in total 17,210,368 possible combinations can be generated. 
However, it is noted in our experiments that the different size and shape of patients will lead to bad combinations, for example a very large liver is matched with a patient with a small liver and the liver region will reach outside of the body region. 
In order to keep only the reasonable combination, a filter $f_s$ is created to filter out bad plans based on the normalized size of the anatomical structures 
\begin{equation}
    r_s(M_i, M_0) = |\frac{\sum M_i - \sum M_0}{\sum M_0}|
\end{equation}
where $M_0$ and $M_i$ are the binary mask of the base organ and target organ. 
We assume that human organs will share similar shapes and thus organs with similar size will fit other patients, so the combination of organs are restrained by $r_s$.
In our experiments, we restrain the combination by $r_s<0.02$ and for the case of CT-ORG dataset, the number of total combinations decreases to 1,545 after planning.

\subsection{Object transplant}

For each plan, the component organs are selected for the background but the spatial location in the input image usually will not match the output image. In \anatomix, we transplant the input organs onto the background image by spatial shift.
The offset $(dx,dy,dz)$ is calculated using the binary mask 
\begin{equation}
    M_i^*=\argmax_{d}  \sum M_b \cdot M_i(dx,dy,dz),
\end{equation}
where $M_i$ is the binary mask of the organ $i$ and $M_i(dx,dy,dz)$ shifts $M_i$ by $(dx,dy,dz)$. After shifting, the input mask is assured to have the biggest overlap with the background mask. And the augmented organs mask is
\begin{equation}
    M^* = \{M_i^*|0<i<N_t\}
\end{equation}
and the output image is 
\begin{equation}
    I^*=I_b + \sum_{0<i<N_t} I^*_i\cdot M_i^*,
\end{equation}
where $I_b$ is the background image and $I_i^*$ is the input image of organ $i$ shifted by $(dx,dy,dz)$.


\subsection{Data}
\anatomix is evaluated on the CT-ORG dataset.\cite{rister2020ctorg} CT-ORG contains 120 throat-abdominal CT volumes with segmentation of lungs, liver, kidneys, bladder and bones. After filtering images with incorrect pixel-dimension, 108 images are used in the following experiments. In particular, 28 images are selected for training and 80 images for test, in order to maximally keep the robustness of the test dataset.


\subsection{Experimental Setting}
Three experiments are designed to investigate the \anatomix on different amount of training data, where 10, 20 and 28 images are used for training the segmentation model. 
For each experiment, we apply \anatomix offline to augment 500 images from the base dataset, namely the atmx dataset, and then a plain U-Net is trained on each dataset, in comparison with the baseline model where no data augmentation is applied.
The dice score of each organ and the average dice are used for evaluation of segmentation performance. All experiments are done on 1 Nvidia 1080 (8GB).

\section{Results}

\begin{figure}[ht]
    \centering
    \includegraphics[width=1.0\textwidth]{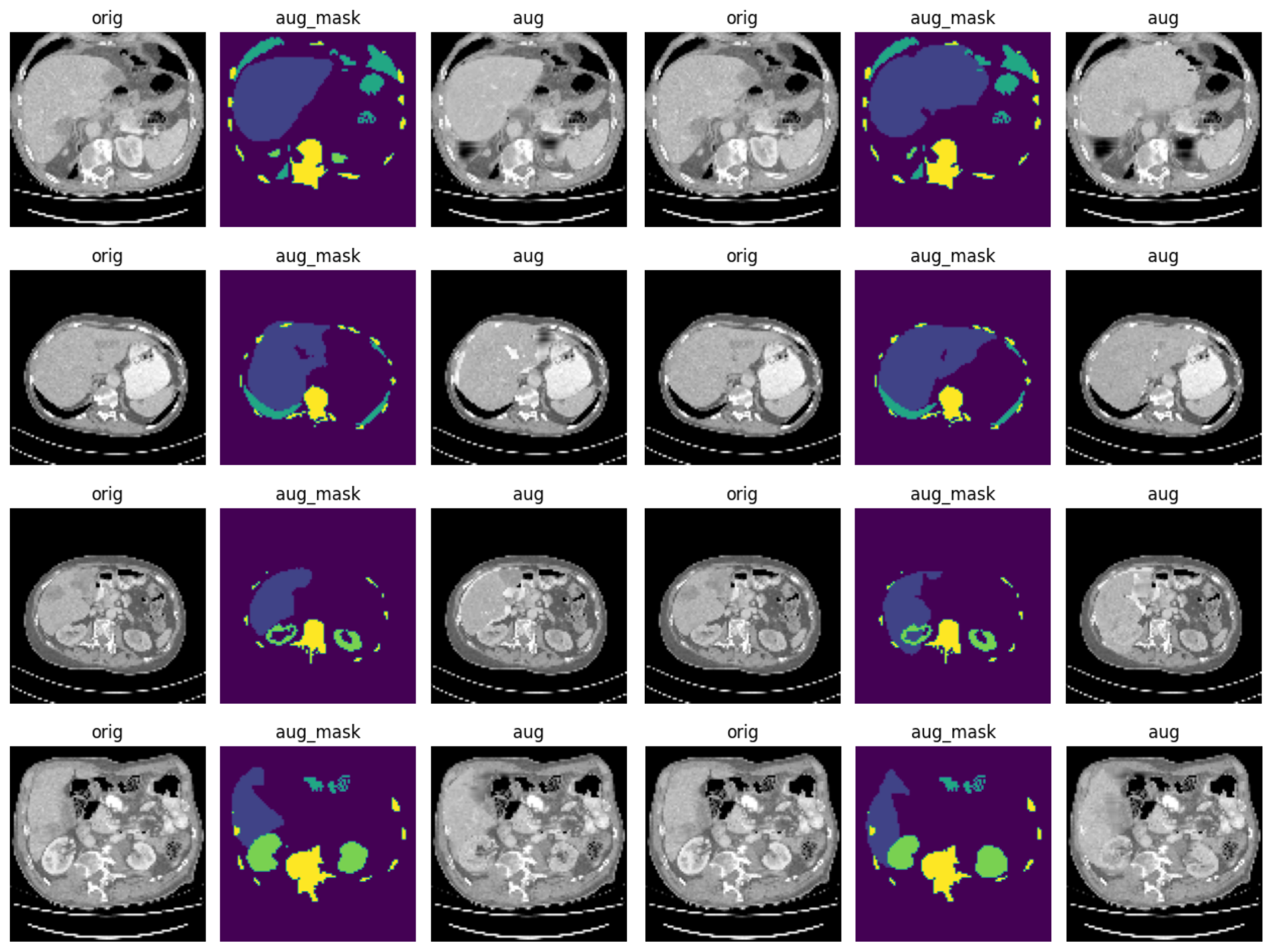}
    \caption{Some example slices of the outputs of \anatomix.}
    \label{fig:aug_res}
\end{figure}
Some example slices of the outputs of \anatomix are shown in Fig. \ref{fig:aug_res}. As shown in Fig. \ref{fig:aug_res}, \anatomix can produce the CT volumes with correct organ location and similar organ size. 
From the initial results in Table. \ref{tab:res:ctorg}, \anatomix can lead to improvement of average dice by 3.2\% and 1.7\% when 10 and 20 images are used for image augmentation. When 28 images are used for training, no improvement is observed using AnatoMix.

\begin{table}[ht]
\centering
\caption{For each experiment, \anatomix creates 500 augmented images from the corresponding 10, 20, 28 training data. The dice score of each organ is then evaluated.}
\begin{tabular}{|l|l|l|l|l|l|}
\hline
  & avg   & liver & lung  & kidney & bone  \\ \hline
base10 & 0.717 &\textbf{0.792} & \textbf{0.948} & 0.502  & 0.628 \\
atmx10 & \textbf{0.740} & 0.787 & 0.944 & \textbf{0.576}  & \textbf{0.651}\\ \hline
base20 & 0.748 &\textbf{0.809} & 0.940 & 0.561  & 0.681 \\
atmx20 & \textbf{0.761} & 0.803 & \textbf{0.955} & \textbf{0.598 } & \textbf{0.689} \\ \hline
base28 &\textbf{0.793} & \textbf{0.855} & 0.961 & 0.617  & \textbf{0.737} \\
atmx28 & 0.787 & 0.827 & 0.961 & \textbf{0.659}  & 0.703 \\ \hline
\end{tabular}
\label{tab:res:ctorg}
\end{table}

\section{Discussion}
In this work, we propose \anatomix, an object-level data augmentation method to improve the generalizability of multi-organ segmentation datasets, and investigate its performance on the CT-ORG dataset for initial evaluation. It is shown in our experiments, that \anatomix is able to increase the segmentation performance in some cases. 

During our experiments, it is also noticed that \anatomix can degrade the segmentation performance when the size of the training dataset is increased. Possible reason would be because of the intrinsic annotation errors in the CT-ORG dataset. Future research is still undergoing to test \anatomix on other multi-organ segmentation datasets.

\clearpage

\bibliographystyle{splncs04}
\bibliography{citation}
\end{document}